\documentclass [10pt,compsocconf]{IEEEtran}
\usepackage{algorithm,amsbsy,amsmath,amssymb,epsfig,bbm,mathrsfs,multirow,amsthm}
\usepackage{algpseudocode}
\usepackage{subcaption,float}

\usepackage{mathtools} 
\usepackage{tabularx}
\usepackage{cite}
\makeatletter
\newcommand{\multiline}[1]{%
	\begin{tabularx}{\dimexpr\linewidth-\ALG@thistlm}[t]{@{}X@{}}
		#1
	\end{tabularx}
}
\makeatother

\begin{document}
	\title{Fast or Slow: An Autonomous Speed Control Approach for UAV-assisted IoT Data Collection Networks\\ }
	\author{\IEEEauthorblockN{Nam H. Chu, Dinh Thai Hoang, Diep N. Nguyen, Nguyen Van Huynh, and Eryk Dutkiewicz \\}
		School of Electrical and Data Engineering, University of Technology Sydney, Australia\\
		\vspace{-5mm}}

	\maketitle
	\thispagestyle{empty}
	\begin{abstract}
		Unmanned Aerial Vehicles (UAVs) have been emerging as an effective solution for IoT data collection networks thanks to their outstanding flexibility, mobility, and low operation costs. 
		However, due to the limited energy and uncertainty from the data collection process, speed control is one of the most important factors to optimize the energy usage efficiency and performance for UAV collectors. 
		This work aims to develop a novel autonomous speed control approach to address this issue. 
		To that end, we first formulate the dynamic speed control task of a UAV as a Markov decision process taking 
		into account its energy status and location.	
		In this way, the Q-learning algorithm can be adopted to obtain the optimal speed control policy for the UAV. 
		To further improve the system performance, we develop an highly-effective deep dueling double Q-learning algorithm utilizing outstanding features of the deep neural networks as well as advanced dueling architecture to quickly stabilize the learning process and obtain the optimal policy. 
		Through simulation results, we show that our proposed solution can achieve up to 40\% greater performance compared with other conventional methods. 
		Importantly, the simulation results also reveal significant impacts of UAV's energy and charging time
		on the system performance. 		
	\end{abstract}
	
	{\it Keywords-} IoT, UAV, data collection, speed control, deep Q-learning, MDP, and deep dueling. 
	
	\section{Introduction}
	\label{sec:Introduction}
	The development of IoT systems has been explosively evolving over the past years to support
	various aspects of our daily lives. 
	It was forecasted that by 2023, IoT connections would represent 50\% of mobile connections (equivalent to 14.7 billion connections), up from 33\% in 2018 \cite{Cisco}.
	However, the expansion of the network scale raises numerous challenges for network operators and service providers. 
	First, the IoT devices are often distributed sporadically in large areas, making it inefficient to deploy conventional wireless access points (e.g., Wi-Fi access points) to collect data from all IoT devices.
	Second, in some cases, it is impossible to collect data from IoT devices by using conventional access points, e.g., when IoT devices are attached to the bridges, outside of the buildings, or even on top of trees.  
	Another approach is using cellular base stations to collect data from IoT devices.
	However, due to the limited energy and communication capability, IoT nodes cannot transmit data in a long distance. Thus, effective solutions for IoT data collection networks are in urgent need. 
	
	Recently, UAVs have been introduced as a very promising solution to address the aforementioned challenges.
	Specifically, with aerial superiority, UAVs can act as on-demand access points that can provide good line-of-sight (LoS) paths for IoT devices, resulting in better communications and quality of service (QoS) compared to other conventional methods, especially in latency- and data rate-sensitive IoT applications~\cite{Wang2019Joint}. 
	Moreover, in remote regions where terrestrial infrastructures are not available, the deployment of UAVs is much more economic than conventional methods, e.g., using high-cost satellite connections or deploying long-range ground broadcasting stations. 
	More importantly, UAVs can be quickly deployed for emergency situations where the existing communication infrastructure is damaged and unable to collect data from IoT devices~\cite{Liu2020Distributed}.		
	However, there are several challenges which are impeding the development of UAV's applications in IoT data collection networks. 	
	First, unlike conventional data collection methods, e.g., using fixed wireless access points, UAVs have limited energy supply. 
	Specifically, UAVs are usually equipped with batteries for operations, and when the UAVs run out of batteries, they have to go to the charging stations to charge or replace the batteries. 		
	Thus, efficient energy usage is the critical step to achieve a high performance for the UAV-assisted IoT data collection networks. 		
	Second, UAVs are usually moving to collect data, while IoT devices are stationarily distributed in different areas and data generated from them are random depending on sensing information obtained from surrounding environments.	
	Thus, controlling UAVs' operations in different areas to maximize data collection efficiency is a big challenge in UAV-assisted IoT data collection networks. 
	
	To address the above problems, speed control is an effective solution to maximize data collection efficiency. 
	Specifically, due to the fact that different locations may have different numbers of IoT devices and different amount of sensing data, the UAVs need to control their speeds over different locations to maximize data collection efficiency. 
	For example, in some places where there are many IoT devices, the UAVs may want to fly at a low speed to increase opportunities to collect data. 
	In contrast, at other places where less or even no IoT devices present, the UAVs can increase its flying speed to avoid missing opportunities from other active places. 
	Furthermore, it can be observed that flying at a low speed will consume less energy than flying at a high speed. 
	Thus, by controlling the UAVs' speed appropriately, we can not only maximize opportunities to collect data, but also save more energy, and thereby improving the overall system performance for UAV-assisted IoT data collection networks.	

	In the literature, several studies consider UAVs' applications for IoT data collection networks~ \cite{Gong2018JSAC,Pan2018Sensor,Lin2019JIoT}.
	In \cite{Gong2018JSAC}, the authors aimed to find the optimal speed for a UAV to minimize the overall flight time for the data collection task. 
		However, this work requires complete information from the IoT devices in advance and does not consider impacts of energy as well as charging process for the UAV during its data collection task. 		
		In \cite{Pan2018Sensor} and \cite{Lin2019JIoT}, the authors proposed a dynamic speed control algorithm that can adaptively adjust the UAV's speed according to the density of ground devices to maximize the data collection efficiency. 
		However, similar to \cite{Gong2018JSAC}, these works also assume that the information about data collected from IoT devices is known in advance and both of them do not consider impacts of energy consumption and charging process during the data collection process. 
		It is important to note that, as UAVs have limited energy, energy charging is an important process which cannot be ignored.
		In addition, data generated from IoT devices depends on not only IoT density, but also information collected from surrounding environment. 
		Thus, effective solutions to these problems need to be further investigated.  		
	
	In this paper, we introduce an effective framework to optimize the performance for a UAV-assisted IoT data collection system through dynamically controlling the UAV's speed. 	
	In particular, we consider a UAV flying over an area with a predefined trajectory to collect sensing data from IoT devices distributed in the considered area. 		
	To maximize the UAV's performance under the uncertainty of the data collection process and the limited capacity of UAV's energy storage, we formulate the UAV's dynamic speed control task as a Markov decision process (MDP). 
	In this way, a Q-learning algorithm can be adopted to help the UAV to find the optimal speed control policy without requiring information about data generation statistic from IoT devices and charging time in advance. 		
	Due to the high-complexity of the considered dynamic optimization problem, we develop an advanced reinforcement learning algorithm, called Deep Dueling Double Q-Learning (D3QL) to quickly find the optimal policy for the UAV. 		
	The key idea of this algorithm is based on recent advances of the deep dueling neural network architecture~\cite{Wang2016Dueling} to separately and simultaneously estimate the values of states and advantages of actions. 		
	Simulation results show that our proposed algorithm can achieve much stable and superior performance than those of conventional approaches (e.g., up to 40\% greater performance compared with a fixed-speed policy).		
	Furthermore, the results also demonstrate the efficiency of our proposed solution in terms of energy and throughput through optimizing flying speed and serving time. 		
	To the best of our knowledge, this is the first work in the literature studying an autonomous flying approach taking energy limitation, dynamic of data collection process, and impact of charging process, into considerations.					
	\begin{figure}[!]
		\begin{center}
			$\begin{array}{c}
				\includegraphics[width=0.8\linewidth]{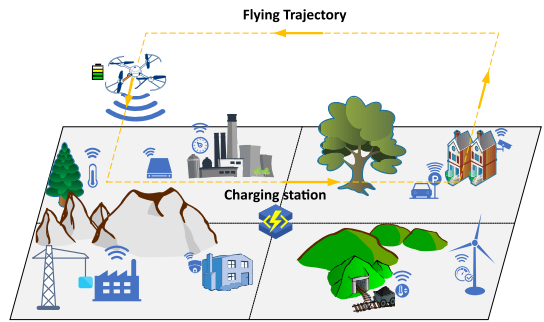}\\ 
			\end{array}$	
			\caption{System model.}
			\label{fig1}
		\end{center}
	\end{figure}
	\section{System Model}
	\label{sec:sysmodel}		
	We consider an IoT data collection system assisted by a UAV, as depicted in Fig. \ref{fig1}.
	In particular, a set of IoT nodes are deployed to perform diverse sensing tasks (e.g., temperature and humidity) over a  considered area which can be partitioned into $N$ cells.
	Due to the fact that some sub-areas may need more attention than others depending on specific IoT applications and tasks, the IoT devices may be unevenly distributed over these cells. 
	The UAV acts as a flying data collector that obtains data from the IoT nodes.
	We assume that time is slotted (as in \cite{Pan2018Sensor,Lin2019JIoT}).
	A time slot is divided into broadcasting and transmission periods.
	At the beginning of a time slot, the UAV broadcasts a message to notify ground devices that it is ready to serve.
	Upon receiving the message, IoT nodes in the communication range transmit their data to the UAV.
	In each cell, we assume that the OFDMA technique is employed for the uplink, i.e., from IoT nodes to the UAV, and the UAV uses another dedicated channel for the downlink, i.e., from the UAV to IoT devices \cite{Wu2018OFDMA}. 
	The probability of receiving a data packet in one time slot in cell $n$ is denoted by $p_n$.
	Since the numbers of IoT devices in each cell are different and their applications are diverse, the probabilities of receiving data vary over these cells. 
	As a result, the UAV may receive more data when flying on the cell with a high value of $p_n$.
	However, these probabilities are highly dynamic and unknown by the UAV in advance.
	Therefore, the UAV needs to learn this information to maximize its data collection efficiency.	
	
	In this work, we assume that the UAV follows a predefined trajectory designed (as in \cite{Gong2018JSAC,Lin2019JIoT,Pan2018Sensor}) to ensure that it can sweep through all nodes in each round.	
	The UAV can fly with different speeds in different locations, but the speed in a time slot is assumed to be constant.
	In addition, unlike \cite{Gong2018JSAC,Pan2018Sensor,Lin2019JIoT}, in this paper, we consider a practical scenario in which the UAV is equipped with a limited energy storage with maximum $E$ energy units. 		
	When the energy level is below a predefined threshold, the UAV will fly to the nearest charging station to replace the battery. 		
	After that, the UAV will return to its trajectory to continue collecting data from IoT devices. 
	Note that the charging time is usually unknown by the UAV in advance and it is dynamic depending on the current location of the UAV and the charging stations. 		
	Thus, in the next sections, we introduce an advanced reinforcement learning algorithm, i.e., D3QL, which can quickly obtain the optimal speed control policy for the UAV given the uncertainty of data collection process and limited energy of UAV. 
	
	\section{Optimal Speed Control Formulation}
	\label{sec:LA}
	In this section, we adopt an MDP framework to formulate the dynamic speed control problem for the UAV. 
	This framework can help the UAV to choose the best speed based on its current location and energy level to maximize its average long-term reward without requiring information about charging time and data collection statistic of IoT devices in advance.  	
	\subsection{State Space}
	The system state is defined based on the current location and the energy level of the UAV (denoted by $e$). 
	The location of UAV can be determined based on two factors, i.e., the current cell number, i.e., $c$, and its current location in this cell, i.e., $l$. 
	Thus, the state space of the UAV is defined by:
	\begin{equation}
		\begin{aligned}
			\mathcal{S}& = \Big\{ ({\mathcal{C}}, {\mathcal{L}}, {\mathcal{E}}): {\mathcal{L}} \in \{0,\ldots,c, \ldots, C\}; \\
			& {\mathcal{L}} \in \{0,\ldots,l,\ldots, L \};
			\quad \text{and} \phantom{5} {\mathcal{E}} \in \{0,\ldots,e,\ldots, E \}  \Big\},
		\end{aligned}
	\end{equation}
	where $C$ and $L$ are the maximum numbers of cells and locations in each cell, respectively. 
	$E$ is the UAV's maximum energy capacity.
	In this way, the system state can be expressed as a tuple $s = (c,l,e) \in \mathcal{S}$.
	In addition, due to the charging process, we need to define a special state, i.e., $s=(-1,-1,-1)$. 
	The system will go to this state only when the current energy level is under a predefined threshold, i.e., when the current state is $s=(c,l,0)$. 	
	Then, after the charging period, the UAV will come back to the last location before it goes to the charging station with full energy, i.e., $s=(c,l,E)$. In this way, our system process will be continuous and without failing to a terminating state.
	\subsection{Action Space}
	At each time slot, the UAV can choose to fly with a speed $a$ selected from the set $\mathcal{A} = \{1,\ldots,A\}$ where $A$ is the highest speed level that the UAV can choose from.
	
	\subsection{Reward Function}
	In this work, we aim to optimize the system performance through tradeoffs between the data collection efficiency and energy usage efficiency. 
	The data collection efficiency can be defined as the number of data packets that the UAV can collect over a time slot and the UAV working status.  
	For example, in a considered time slot, if the UAV is working, i.e., flying to collect data, the UAV will receive a reward $\Omega$. 
	However, if the UAV is at the charging state, it will receive a reward of zero. 
	This design reward function is to \emph{``push''} the UAV spending more time for collecting data instead of lying on the charging stations for charging the battery. 
	In addition, we add a cost for consuming energy if the UAV is flying. 
	Obviously, the energy consumption per time slot will depend largely on the chosen speed. 
	Specifically, the higher speed the UAV chooses to fly, the more energy it consumes. 
	If we denote $m^a_t$ as the energy consumption when the UAV flies at speed $a$ in time slot $t$, the immediate reward function for the UAV can be defined by: 		
	\begin{align}
		&r_t(s_t,a_t) = \left\{\,	\begin{array}{ll} \Omega +	w_1 d^s_t - w_ 2 m^a_t ,&	\mbox{if the UAV is working,}	\\
			\,0	,																	&	\mbox{otherwise,}
		\end{array}	\right.
	\end{align}	
	where, $d^s_t$ is the number of data packets the UAV receives at the current state $s_t$. Moreover, $w_1$ and $w_2$ are the weights to trade-off between the amount of collected data and energy consumption. 
	In this way, the reward function can capture not only data collection efficiency, but also the energy consumption efficiency for the UAV.
	
	In this paper, our objective is to find an optimal policy $\pi^*$, i.e., a mapping from the state space to the action space \mbox{$\pi^*:\mathcal{S} \rightarrow \mathcal{A}$}, to maximize the long-term average reward function defined as follows:
	\begin{eqnarray} 
		\label{eq:average_reward}
		\max_\pi \quad	{\mathcal{R}}(\pi)	=	\lim_{T \rightarrow \infty} \frac{1}{T} \sum_{t=1}^{T} {\mathbb{E}} \left( r_t (s_t, \pi(s_t)) \right),	\label{eq:cmdp_obj}
	\end{eqnarray}
	where ${\mathcal{R}}(\pi)$ is the long-term average reward that UAV receives under the policy $\pi$, and $r_t (s_t, \pi(s_t))$ is the immediate reward under policy $\pi$ at time $t$. The optimal policy $\pi^*$ will allow the UAV to make the optimal decision dynamically based on its current state, i.e., the cell, location, and remain energy.
	
	\section{Q-Learning Algorithm}
	\label{sec:Qlearning}
	To obtain the optimal policy for the UAV under uncertainty of the data collection and energy charging processes, Q-learning algorithm \cite{Watkins1992QLearning} is adopted in this work. 
	The main reason for using this algorithm is due to its outstanding features. 
	Specifically, this algorithm can help the UAV to obtain the optimal policy through gradually learning from surrounding environment (e.g., how often the UAV can obtain data packets in different areas as well as how long it often needs to replace the battery) during its service time. 
	To that end, at the beginning, we assume that the UAV starts following a policy $\pi$ from state $s$, where $s \in \mathcal{S}$. 
	Then, the state-value function of state $s$ under policy $\pi$, which specifies how good to be in this state, can be determined as \cite{Sutton1998Reinforcement}
	\begin{equation}
		\begin{aligned}
			\mathcal{V}^\pi(s) &\triangleq \mathbb{E}_\pi \Big [ \sum_{t=0}^{\infty} \gamma^t r_t(s_t, a_t)~\big|~s_0=s\Big ]\\
		\end{aligned}
	\end{equation}
	where $r_t$ is the immediate reward achieved by taking action $a_t$ at state $s_t$, and $\gamma \in [0,1)$ is the discount factor that represents the significance of long-term rewards~\cite{Watkins1992QLearning}. 
	Next, we define the action-value function under policy $\pi$ for action $a$ at state $s$, named Q-function, as follows:
	\begin{equation}
		\label{eq:q-function}
		\begin{aligned}
			\mathcal{Q}^\pi(s,a) &\triangleq \mathbb{E}_\pi \Big [ \sum_{t=0}^{\infty} \gamma^t r_t(s_t, a_t)~\big|~s_0=s, a_0=a\Big ].
		\end{aligned}
	\end{equation}
	The Q-learning algorithm maintains a table to learns the optimal value of Q-function, denoted by $\mathcal{Q}^*(s,a)$ through iteratively updating this table. 	
	After the UAV takes action $a_t$ at time $t$, it observes reward $r_t$ and next-state $s_{t+1}$. 
	Then, the Q-function is updated by the temporal difference (TD), which is the different between target Q-value, i.e., $Y_t=r_t(s_t, a_t) + \gamma\max_{a_{t+1}} \mathcal{Q}_t(s_{t+1}, a_{t+1})$, and the current estimated Q-value, i.e., $\mathcal{Q}_t(s_t,a_{s_t})$, as follows: 	
	\begin{equation}
		\label{Eq:updateQfunction}
		\begin{aligned}
			\mathcal{Q}_{t}(s_t,a_t) \leftarrow &\mathcal{Q}_t(s_t,a_t) + 
			\beta_t\Big [ Y_t - \mathcal{Q}_t(s_t,a_{s_t})\Big ],
		\end{aligned}
	\end{equation}
	where $\beta_t$ is the learning rate that demonstrates the impact of new information, i.e., temporal difference. The learning rate $\beta_t$ can be a constant or adaptively changed during the learning process. 
	The Q-learning algorithm is proved to be converged to the optimal policy with probability one if the learning rate $\beta_t$ satisfies \eqref{Eq:rules} and Q-function is updated by \eqref{Eq:updateQfunction}~\cite{Watkins1992QLearning}.
	\begin{equation}
		\label{Eq:rules}
		\beta_t \in [0,1), ~\sum_{t=1}^{\infty}\beta_t = \infty, \mbox{ and } \sum_{t=1}^{\infty} ( \beta_t  )^{2} < \infty.
	\end{equation}
	However, the Q-learning algorithm usually requires a long time to find the optimal policy for the UAV due to a high dynamic of the interacting environment caused by the uncertainty of data collection and energy charging processes. 	
	Therefore, in the next section, we develop an effective Deep Dueling Double Q-learning (D3QL) algorithm to address this problem.
	\section{Deep Dueling Double Q-learning}
	Recently, deep Q-learning algorithms with experience replay have been introduced and made a breakthrough in solving complex problems in practice~\cite{Mnih2015Human}. 
	Using nonlinear estimators, expressed by deep neural networks, for approximating the Q-value, it can achieve results of playing Atari games comparable to that of humans.
	To further improve the stability of deep Q-learning, the authors in~\cite{Wang2016Dueling} introduced a novel dueling network architecture. 
	In this architecture, the network is spitted into value and advantage streams to represent
	the state-value function $\mathcal{V}^\pi(s)$ and advantage function $\mathcal{O}^\pi(s,a)$, respectively.
	The advantage function under policy $\pi$ is defined by 
	\begin{equation}
		\label{eq:Dueling1}
		\mathcal{O}^\pi(s,a) \triangleq	\mathcal{Q}^\pi(s,a) - \mathcal{V}^\pi(s).
	\end{equation}
	The value function represents the quality of a given state, and the advantage function expresses the importance of an action compared with others. 
	At the output layer of the network, two streams are combined to estimate the value of Q-function by
	\begin{equation}
		\label{eq:Dueling2}
		\mathcal{Q}(s,a;\phi, \zeta) = \mathcal{V}(s;\phi)
		+ \mathcal{O}(s,a;\zeta),
	\end{equation}
	where $\phi$ and $\zeta$ are parameters of the value stream and advantage stream, respectively. 
	Because $\mathcal{V}$ and $\mathcal{O}$	cannot be recovered from $\mathcal{Q}$, \eqref{eq:Dueling2} is unidentifiable~\cite{Wang2016Dueling}.
	This issue is resolved by subtracting the average of the output from the advantage stream as follows:
	\begin{equation}
		\label{eq:Dueling3}
		\begin{aligned}
			\mathcal{Q}(s,a; \phi, \zeta) = \mathcal{V}(s;\phi) + \Big(\mathcal{O}(s,a;\zeta)
			-\frac{1}{|\mathcal{O}|}\sum_{a' \in \mathcal{A}}\mathcal{O}(s,a';\zeta) \Big).
		\end{aligned}
	\end{equation}
	In this way, the dueling architecture is more stable in estimating values of Q-function.
	\begin{algorithm}[t]
		\caption{The D3QL Algorithm}
		\label{alg:deepqlearning}
		\begin{algorithmic}[1]
			\State Initialize replay buffer $\mathbf{B}$ with capacity $\mathcal{B}$.
			\State Initialize Q-network $\mathcal{Q}$ with random parameters $\phi ~\text{and}~ \zeta$ for the value stream and advantage stream, respectively.
			\State Initialize target Q-network $\hat{\mathcal{Q}}$ as a copy of the Q-network with parameters $\phi^-=\phi ~\text{and}~ \zeta^-=\zeta$.
			\For{\textit{step = 1 to T}}
			\State \multiline{Choose a random action $a_t$ with probability $\epsilon$ , otherwise select $a_t=\arg \max \mathcal{Q}^*(s_t, a_t; \phi, \zeta)$.}
			\State \multiline{Perform $a_t$, obtain reward $r_t$ and next state $s_{t+1}$.}
			\State Store transition $(s_t, a_t, r_t, s_{t+1})$ in replay buffer $\mathbf{B}$.
			\State \multiline{Sample random mini-batch of transitions $(s_k, a_k, r_k, s_{k+1})$ from $\mathbf{B}$.}
			\State \multiline{Combine the value function and the advantage function by \eqref{eq:Dueling3}.}
			\State \multiline{Calculate target Q-value $Y_t$ by \eqref{eq:targetDQN}.}
			\State \multiline{Perform a gradient descent step with respect to Q-network parameters on $\big(Y_k-\mathcal{Q}(s_k, a_k; \phi, \zeta)\big)^2$.}
			\State Every $I$ steps set  $\hat{\mathcal{Q}}= \mathcal{Q}$.
			\EndFor
		\end{algorithmic}
	\end{algorithm}

	However, both Q-learning and deep Q-learning algorithms might not perform well in stochastic MDP since it usually overestimates the value of Q-function, i.e., $\mathcal{Q}(s,a)$~\cite{Hasselt2016}.
	This issue negatively impacts optimal policies and even leads to sub-optimal policies if they are not uniformly distributed over the states\cite{Thrun1993}.
	Double DQN is thus proposed to address this issue by using two estimators, one to select an action and another to calculate the Q-value \cite{Hasselt2016}.
	Therefore, we develop the D3QL algorithm which can leverage the advantages of both deep dueling network architecture and double DQN to reduce the overestimation of deep Q-learning and improve the stability of the learning process, and thereby quickly obtaining the optimal policy for the UAV.	
	The details of the D3QL algorithm are presented in Algorithm~\ref{alg:deepqlearning}. 
	Specifically, the training phase of D3QL takes $T$ steps. 
	For state $s_t$ at step $t$, the algorithm first chooses an action according to $\epsilon$-greedy policy.
	Then, it observes reward $r_t$ and next state $s_{t+1}$ at the end of the time slot.
	These observations are then stored in a buffer $\mathbf{B}$.	
	Instead of using only the current information, i.e., a transition \mbox{$(s_t, a_t, r_t, s_{t+1})$}, a mini-batch of transitions is sampled uniformly at random, \mbox{$(s,a,r,s')\sim U(\mathbf{B})$}, to feed the neural network. 
	In this way, the algorithm leverages its experiences and breaks the correlation between consecutive transitions to reduce the variance of the updates.
	In this problem, the input layer has three features representing the UAV's state dimension, i.e., cell number, location in cell, and the energy level. 
	
	The overestimations of Q-learning based algorithms can be large because of a max operator when updating the Q-function in \eqref{Eq:updateQfunction} \cite{Hasselt2016}. 
	In D3QL, we handle this problem by using two estimators, both are the deep dueling networks. 
	We call them Q-network and target Q-network, denoted as $\mathcal{Q}(s,a;\phi, \zeta)$ and $\hat{\mathcal{Q}}(s,a; \phi^-, \zeta^-)$, respectively.
	At step $t$, the target Q-value in \eqref{Eq:updateQfunction} becomes 
	\begin{equation}
		\label{eq:targetDQN}
		Y_t = r_t + \gamma \hat{\mathcal{Q}}\big(s_{t+1}, \underset{a}{\operatorname{argmax}}\mathcal{Q}(s_{t+1},a; \phi_t, \zeta_t);\phi^-_t, \zeta^-_t\big).
	\end{equation}
	For short notation, we denote all the parameters of the Q-network and those of the target Q-network at step $t$ by $\theta_t$ and $\theta^-_t$, respectively.
	The Q-network is trained to minimize the temporal difference error, i.e., a gap between the target Q-value $Y_t$ and the predict Q-value $\mathcal{Q}(s,a;\theta_t)$. 
	Thus, the loss function of the Q-network at step $t$ is defined as
	\begin{equation}
		\begin{aligned}
			\label{lossfunction}
			L_t(\theta_t) = \mathbb{E}_{(s,a,r,s')\sim U(\mathbf{B})}\bigg[ \bigg( Y_t
			-\mathcal{Q}(s,a;\theta_t)\bigg)^2\bigg].
		\end{aligned}
	\end{equation}
	To minimize the loss function in \eqref{lossfunction}, we can use Gradient Descent (GD) algorithms.
	In particular, GD solves the problem by iteratively updating the network's parameters $\theta$ as follows:
	\begin{equation}
		\label{eq:GDupdate}
		\theta_{t+1} = \theta_t -\eta_t\nabla L_t(\theta_t),
	\end{equation}
	where $\eta_t>0$ is a step size at step $t$. Due to noises occurring during the sampling, the step size $\eta_t$ needs to be decreased over iterations to ensure the convergence.
	The gradient of the loss function in \eqref{lossfunction} with respect to the neural network's weights is calculated as follows:
	\begin{equation}
		\begin{aligned}
			\label{gradient_loss}
			&\nabla_{\theta_t}L(\theta_t) = \mathbb{E}_{(s,a,r,s')}\bigg[\bigg(Y_t
			- \mathcal{Q}(s,a;\theta_t)\bigg)\nabla_{\theta_t}\mathcal{Q}(s,a;\theta_t)\bigg].
		\end{aligned}
	\end{equation}
	\begin{figure}[!]
		\begin{center}
			$\begin{array}{c}
				\includegraphics[width=0.7\linewidth]{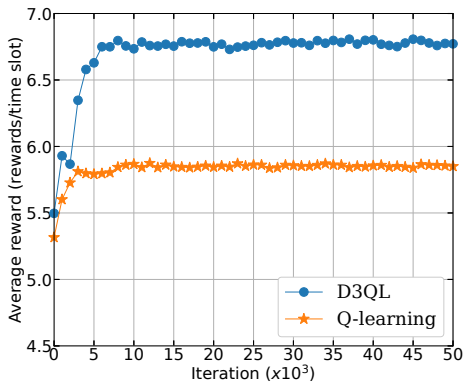}\\ 
			\end{array}$	
			\caption{Convergence rate of D3QL and Q-learning.}
			\label{fig:cvg}
		\end{center}
	\end{figure}	
	\begin{figure}[!]
		\begin{center}
			$\begin{array}{c}
				\includegraphics[width=0.7\linewidth]{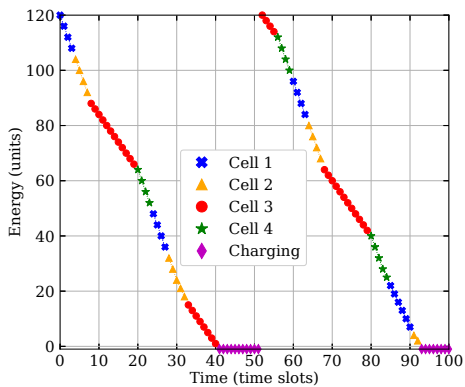}\\ 
			\end{array}$	
			\caption{UAV's optimal policy obtained from D3QL.}
			\label{fig:policy}
		\end{center}
	\end{figure}
	\begin{figure*}[h!]
		\begin{center}
			$\begin{array}{ccc} 
				\includegraphics[width=0.3\linewidth]{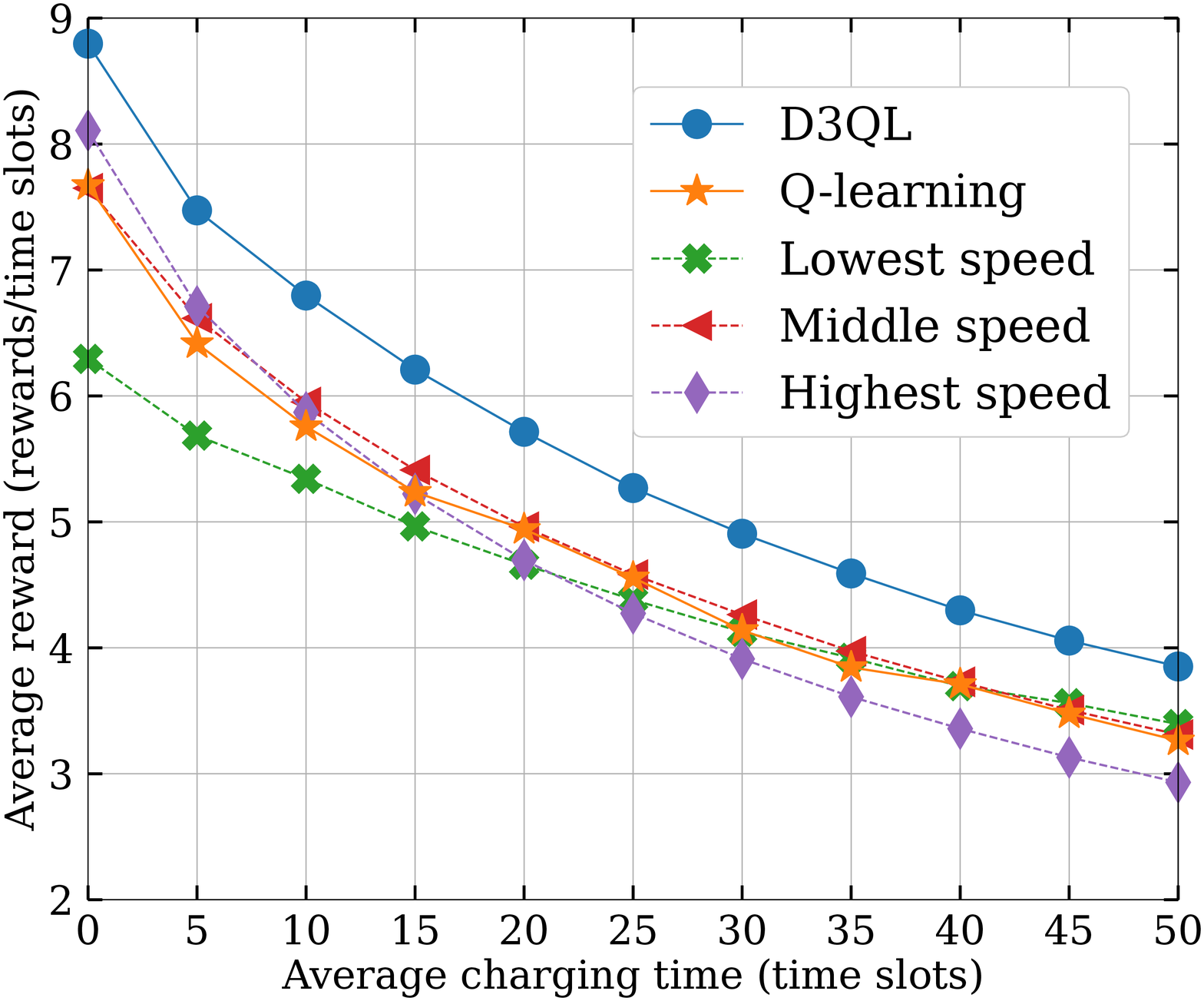} &
				\includegraphics[width=0.3\linewidth]{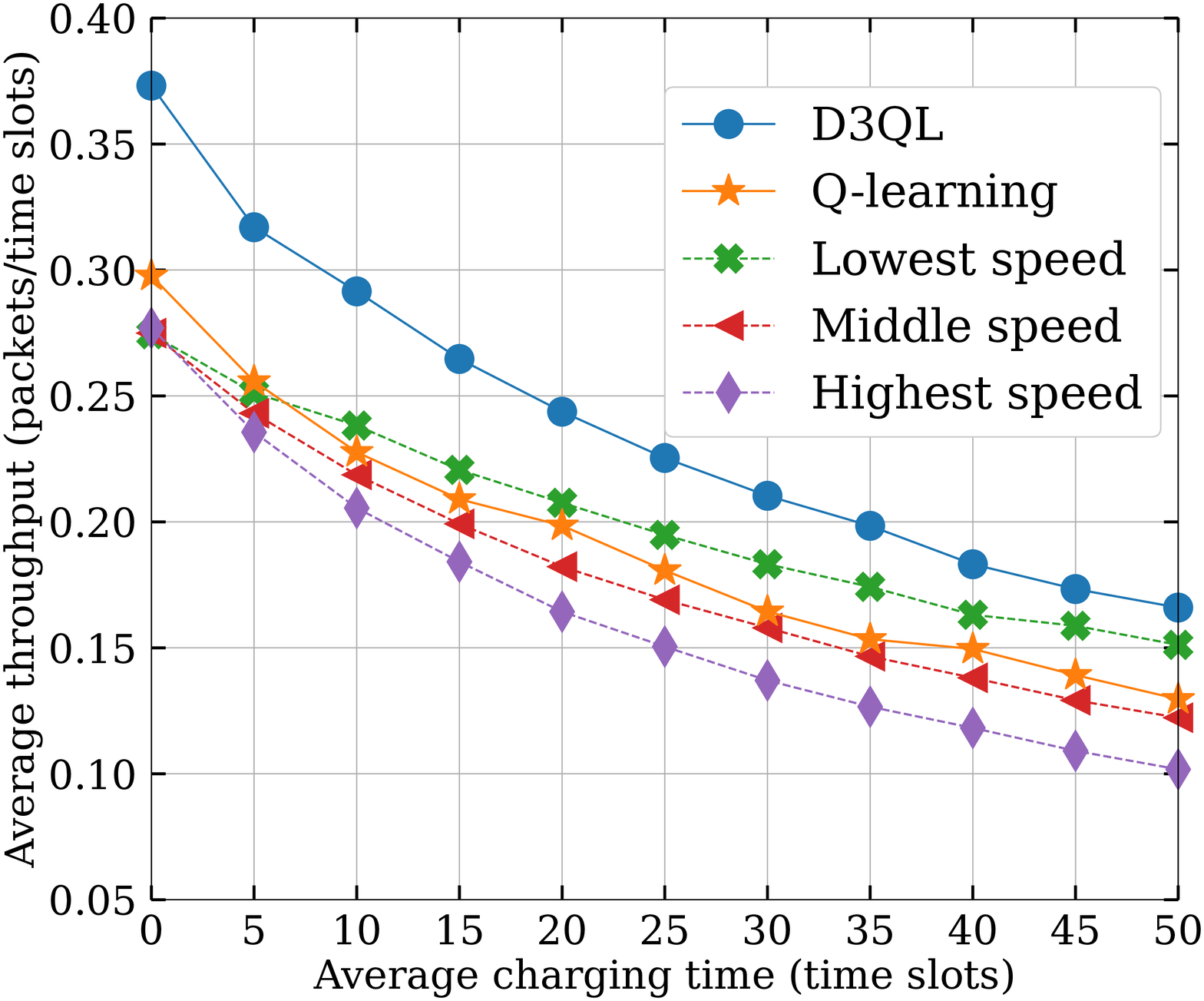} &
				\includegraphics[width=0.3\linewidth]{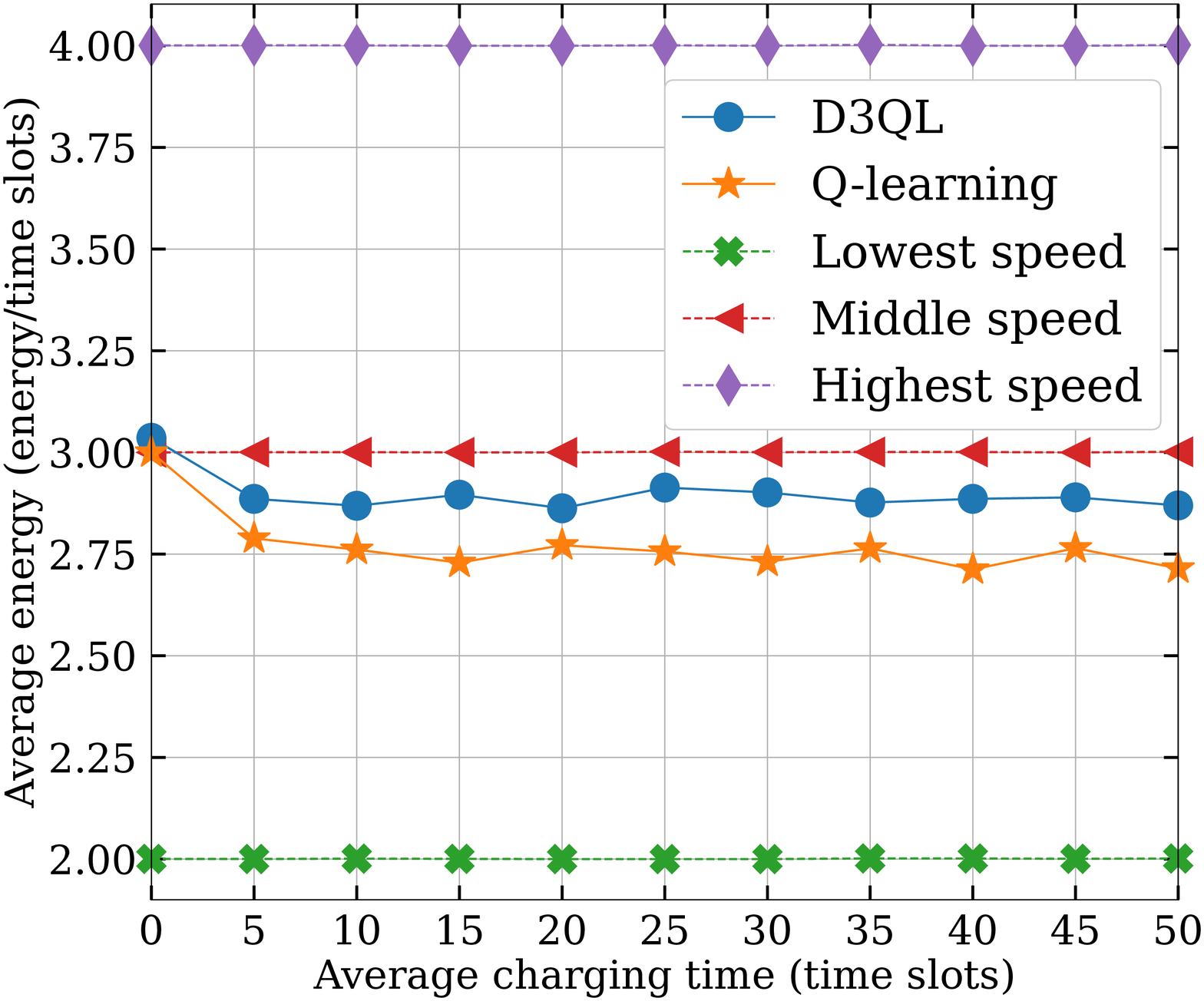} \\
				\text{(a) Average rewards.} &
				\text{(b) Average throughput.}&	
				\text{(c) Average energy.}\\
			\end{array}$
			\caption{Vary charging time.}
			\label{fig:Vary_charging_time}
		\end{center}
	\end{figure*}
	However, GD has to calculate the gradient for all data points for each update which leads to a high computational complexity.
		Therefore, in this work, we adopt Stochastic Gradient Descent (SGD) to achieve faster learning and guarantee convergence \cite{Robbins1951SGD}.	
		Specifically, in each update, SGD only calculates the gradient of a mini-batch sampled uniformly from the memory pool, and thus significantly reducing the computational complexity of the algorithm.
		Note that the Q-network's parameters $\theta$ are updated at every steps, whereas the target Q-network's parameters $\theta^-$ are cloned from $\theta$ at every $I$ steps.
	\section{Performance Evaluation} 
	\subsection{Experiment Setup}	
	In the simulation, we consider a UAV-assisted IoT system covering an area which can be divided into four cells (as illustrated in Fig. \ref{fig1}).
	The distance that UAV travels in each cell is set at $60 m$. 
	The probabilities of receiving data in each time slot in these cells correspond to a coordinated vector $\mathbf{p}=[p_1,p_2,p_3,p_4]$. 
	Note that our proposed algorithms do not require to know $\mathbf{p}$ in advance. 
	They can learn these probabilities through interactions with the environment.
	In IoT data collection networks, UAVs' speeds are usually set at low speeds (e.g., less than 15 m/s) to be able to collect data from IoT devices during its fly [5].
	Thus, we consider three speed levels, i.e., $5m/s, 10m/s,$ and $15m/s$, corresponding to low, medium and high speeds, respectively.
	The corresponding energy consumption for these speed are $2, 3, \text{and}~4$ energy units per time slot, respectively.
	The UAV can use up to $120$ energy units before it has to fly to the nearest charging station to replace the battery.
	The charging time is uncertainty because it depends on the current location of the UAV and its nearest charging station.
	To that end, the charging time is considered as a random variable with mean $z$.
	In the reward function, the value of $\Omega$, $w_1$, and $w_2$ are set to $15, 1$ and $0.5$, respectively.
	Since this paper does not focus on optimizing the deep neural network, the hyperparameters for neural networks are set as typical settings~\cite{Mnih2015Human, Goodfellow2016Deep}. 
	The discount factor $\gamma$ for both reinforcement learning algorithms is $0.9$. The learning rate $\beta$ of the Q-learning algorithm is set to $0.1$. In the $\epsilon$-greedy strategy, $\epsilon$ gradually decays from $1$ to $0.01$. 
	In the simulation, we first investigate the convergence rate of D3QL and Q-learning algorithms.
	We then explore the behavior of proposed algorithms by varying the charging time to evaluate its impact to the system performance.
	This scenario is also to demonstrate the efficiency of the proposed learning algorithms when the UAV can automatically learn from the environment and adapt its optimal policy accordingly.
	For comparisons, we use three deterministic policies, which are flying all the time with the (1) lowest speeds, (2) middle speed, and (3) highest speed. 
	\subsection{Simulation Results}	
	The learning process and convergence rate of the two proposed algorithms can be observed in Fig.~\ref{fig:cvg} when the average charging time $z$ is set to be 10 time slots and the arrival data probabilities in the cells are $\mathbf{p}=[0.1, 0.25, 0.6, 0.15]$. 
	At the beginning, when the proposed algorithms start their learning processes, their average rewards are close to each other, i.e., approximately 5.4.	
	However, only after $10^4$ iterations, the D3QL almost converges and its average reward is more than 15\% greater than that of the Q-learning algorithm.
	This result clearly demonstrates the efficiency of our proposed DQ3L algorithm in dealing with high-complexity systems like what we are considering in this work. 
	
	Next, we investigate the optimal policy of UAV obtained by D3QL in Fig. \ref{fig:policy}.
	Each point represents the UAV's energy level at the beginning of a time slot, and the slope of the line reveals which action is taken in a time slot, e.g., the steeper the slope is, the faster speed is selected. 
	As observed in Fig.~\ref{fig:policy}, the UAV chooses the low speed in a cell with a high probability of receiving data and the high speed in a cell with a low probability of receiving data.
	In particular, given the probabilities $\mathbf{p}=[0.1, 0.25, 0.6, 0.15]$, the fastest speed is selected in cell 1, 2, and 4, whereas the slowest one is chosen in cell 3. 
	More interestingly, the UAV experiences all the speeds in cell 2.
	Specifically, the UAV flies at the highest speed until the energy level decreases to 25 units. 
	After that the middle speed is selected.
	When it nearly runs out of battery, as shown in time slot 91 and 92, the UAV chooses the slowest speed.
	This is stemmed from the fact that the UAV wants to reserve more energy when the current energy level is low, and this result also clearly shows the impact of energy to the optimal decision of the UAV.  
	
	We then evaluate the performance of the proposed learning algorithms when the changing time is varied. 
	In Fig.~\ref{fig:Vary_charging_time}, the policies of both D3QL and Q-learning are obtained after $5\times10^4$ training iterations.
	It can be observed that, for all the policies, as the charging time increases, the average reward and throughput will be decreased as showed in Fig. 4(a) and (b). 
	The main reason is that given a fixed serving time, the more time the UAV spends for charging, the less time the UAV can spend for collecting data. 
	Interestingly, as observed in Fig. 4(a), when the charging time is small, e.g., less than 15 time slots, the highest speed policy can obtain better reward than that of the lowest speed policy. 
	However, when the charging time is large, the highest speed policy obtains the lowest performance compared with other policies. 
	This is due to the fact that, when the UAV requires more charging time, its data collection efficiency is reduced, and thus the UAV needs to reserve more energy by choosing the lowest-speed policy. 
	To the end, our proposed learning algorithm, i.e., D3QL can balance between the data collection and energy consumption efficiency, and thus it always can achieve the best performance compared with other approaches. 	
	
	\section{Summary} 
	This paper has introduced an effective learning approach to automatically control the UAV's speed to maximize performance for the UAV-assisted IoT data collection network. 
	First, to help the UAV to make dynamic speed control decisions based on its current status, e.g., current energy level and location, under the uncertainty of data collection and energy charging processes, the MDP framework has been developed. 
	Then, the Q-learning has been adopted to find the optimal speed control policy for the UAV. 
	To overcome the limitations of Q-learning algorithm, we have then developed the advanced deep reinforcement learning algorithm, called D3QL. 
	This algorithm can benefit from the outstanding advantages of deep dueling network architecture and DQN, thus significantly improving the UAV's learning process. 
	Simulation results then demonstrate the efficiency of our proposed solution, i.e., D3QL, as well as reveal some important information about the impacts of energy and charging processes of UAV on the system performance.
	\bibliographystyle{IEEE}

\end{document}